\documentclass{aa}

\usepackage{amsmath}           
\usepackage{dcolumn}           
\usepackage{graphicx}          
\usepackage{txfonts}           
\usepackage{mathrsfs}          
\usepackage{rotating}          
\usepackage{subfigure}         
\usepackage{lscape}            
\usepackage{afterpage}         
\usepackage{xspace}            
\usepackage{natbib}            
\usepackage{hyperref}          
\usepackage{nicefrac}          
\usepackage{amsfonts}          
\usepackage{mathrsfs}          
\usepackage{textcomp}
\usepackage{chemformula}       
\usepackage{ulem}
\usepackage{xcolor}

\let\ce\ch

\newcommand{\mcl}[3]{\multicolumn{#1}{#2}{#3}}

\newcolumntype{.}{D{.}{.}{-1}}
\newcolumntype{d}[1]{D{.}{.}{#1}}

\makeatletter
\renewcommand*\aa@pageof{, page \thepage{} of \pageref*{LastPage}}
\makeatother

\definecolor{red}{RGB}{255,0,0}

\begin{document} 

\title{Accurate ab initio spectroscopic studies of {promising} interstellar ethanolamine iminic precursors}

\author{D.~Alberton\inst{1,2}
        \and
                N.~Inostroza-Pino\inst{3}
        \and
        Ryan C. Fortenberry\inst{4}
        \and
        V.~Lattanzi\inst{1}
        \and
        C. ~Endres\inst{1}
                \and
                J.~Fuentealba Zamponi\inst{1}
        \and
                P.~Caselli\inst{1}
}

\institute{Center for Astrochemical Studies, Max-Planck-Institut f\"ur extraterrestrische Physik, Gie\ss enbachstr.~1, 85748 Garching, Germany
        \and
        alberton@mpe.mpg.de
        \and
        Universidad Autónoma de Chile, Facultad de Ingeniería, Núcleo de Astroquímica \& Astrofísica, Av. Pedro de Valdivia 425, Providencia, Santiago, Chile, natalia.inostroza@uautonoma.cl
        \and
        Department of Chemistry \& Biochemistry, University of Mississippi, University, MS 38677-1848, USA}

\date{Received \today; accepted --}

\abstract
{The detection of \ch{NH2CH2CH2OH} (ethanolamine) in molecular cloud G+0.693-0.027 adds an additional player to the prebiotic molecules discovered so far in the interstellar medium (ISM).
{As this molecule} might be formed through condensed-phase hydrogenation steps, {detecting one or more of the molecules involved might} help to elucidate the chemical pathway leading to its production.}
{The chemical path involves the formation of four chemical species. In this work, {we 
study the energies of the isomers involved, indicate the best candidates for detection purposes, and provide the distortion constants of the most energetically favoured isomers undetected so far.}}
{We used highly accurate CCSD(T)-F12/cc-pCVTZ-F12 computations to predict the lowest energy isomers as well as their spectroscopic constants, taking corrections for core electron correlation and scalar relativity into account.}
{{  We studied 14} isomers. We find that the lowest energy isomer proposed in previous studies is not the actual minimum. We provide a set of rotational and distortion constants of the two new most stable isomers together with their fundamental vibrational frequencies in order to guide the search for these important astrochemical precursors of prebiotic molecules in the ISM.}
{}

        
\titlerunning{}
        
\maketitle
\section{Introduction} \label{sec:intro}

The desire to understand the molecular origins of life is one of the main drivers for contemporary astrochemical research. In the era of modern space telescopes such as JWST, the amount of data being provided by such observations will undoubtedly enable new and unprecedented discoveries \citep{Gardner_2006_JWST}. While this new telescope is bringing significant new insights for the infrared (IR) Universe, lower-frequency observations including stalwart ground-based radioastronomical rotational spectra will continue to contribute to the field.  Radioastronomy provides insight into where complex chemistry emerges and begins to play its role in the evolution that spans from molecular clouds to habitable planets. Additionally, an increasing number of molecules are employed as molecular tracers, such as CO and \ch{N2H+}, which are used to trace \ch{H2} and \ch{N2}, respectively \citep{Bergin_2001}, and many more tracers might be employed. Such observations are providing a wider picture of the physical evolution of star forming regions, but the chemical processes occurring along the steps that ultimately brought about life are less well understood.

The basic cellular functions for life as we know it are carried out by four classes of molecules. These classes consist of lipids, sugars, amino acids, and nucleic acids. These molecules are in turn mainly formed by hydrogen, carbon, oxygen, nitrogen, and sulphur. Other elements play crucial roles, 
but their abundances are minor when compared to these latter five. 
The idea that these molecules are inherited from the primordial stage of star formation of our Sun is supported by the correlation existing among CHO-, N-, and S-bearing prebiotic molecules detected in the low-mass protostar IRAS 16293-2422 and the bulk composition of the comet 67P/Churyumov-Gerasimenko \footnote{IRAS 16293-2422 is a homologous to our protosolar system, while the comet 67P/Churyumov-Gerasimenko 
represents one of the most primitive material of our Solar System.} \citep{Drozdovskaya_2019}. 
The relation that exists between the abundances of interstellar complex organic molecules \citep[COMs,][]{Herbst-vanDishoeck_2009} observed in comets and in protoplanetary discs \citep{oberg_2015} further supports this idea.
\begin{figure*}[tb!]
 \centering
 \includegraphics[width=1\textwidth]{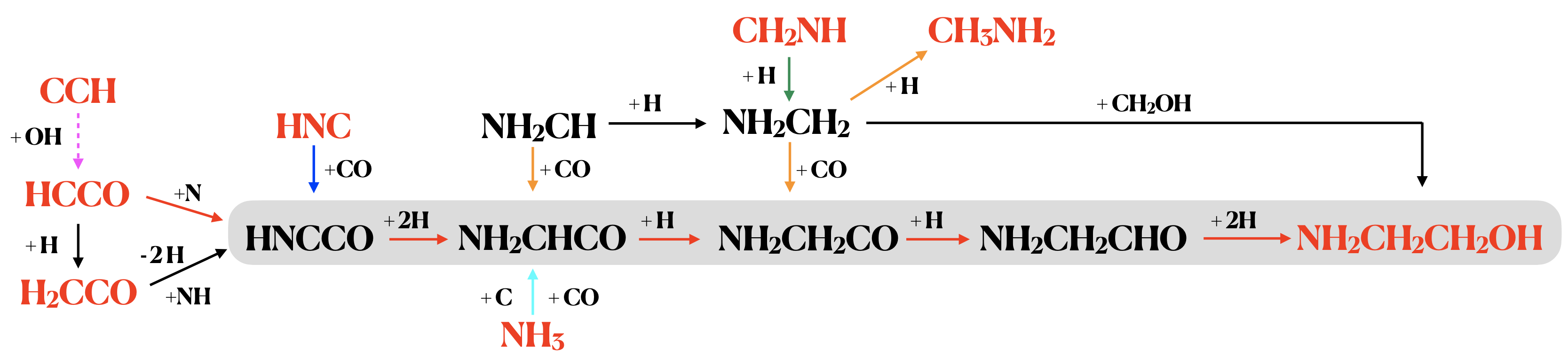}
 \caption{Figure taken from {\citet{rivilla2021a}}. The molecular species in red have been detected towards the G+0.693 molecular cloud. The grey-shaded area corresponds to a hydrogenation chain. The chemical reactions indicated with coloured arrows have been proposed in previous works: magenta \citep{Wakelam_2015}, blue \citep{Kameneva_2017}, orange \citep{Suzuki_2018}, cyan \citep{Krasnokutski_2021}, and green \citep{Suzuki_2018, Ruaud_2015}. The formation routes proposed in {\citet{rivilla2021a}} are shown in black. The solid arrows indicate surface chemistry reactions, dashed arrows denote gas phase chemistry.}
    \label{fig:EtA_form_process}
\end{figure*}
The chemical complexity found in the Murchison and other meteorites \citep{Burton_2012, Pizzarello_2006, Oba_2022} also supports the formation of prebiotic molecules in space, but does not prove causation. As asteroids and planetesimals ---from which meteorites originate--- grew in size, they underwent secondary alteration events. Also known as post-accretion events, these latter provide additional opportunities for chemical reactions, either directly from thermal energy (thermal metamorphism; \citealt{Huss_2006}) or from liquid water (aqueous alteration; \citealt{Brearley_2006}) resulting from melting water ices.
However, the alteration of these bodies makes it difficult to decipher the exact point at which the prebiotic molecules were formed in the ISM, making direct detection the only effective way to determine their presence in space.

Many prebiotic molecules have been detected in the ISM, and both the number and complexity of those that are defined as COMs are sharply increasing \citep{McGuire_2022}.  Precursors of sugars, such as (Z)-1,2-ethenediol \citep[\ch{OHCH=CHOH},][]{rivilla2022a} and glycolaldehyde \citep[\ch{HOCH2CHO},][]{Hollis_2000, Hollis_2004}; nucleobases such as hydroxylamine \citep[NH\textsubscript{2}OH,][]{rivilla2020b} and urea \citep[\ch{CO(NH2)2},][]{Belloche_2019, jimenez2020toward}; and amino acid precursors such as acetonitrile \citep[\ch{CH3CN},][]{belloche2008detection}, vinylamine, ethylamine  \citep[\ch{H2C=CHNH2} and \ch{CH3CH2NH2}, ][]{zeng2021}, and {syn-glycolamide \citep[\ch{NH2C(O)CH2OH},][]{Rivilla_2023ApJ}} have been detected in various astronomical regions. Furthermore, n-propanol \citep[\ch{CH3CH2CH2OH},][]{Jimenez-Serra_2022} and ethanolamine \citep[\ch{NH2CH2CH2OH},][]{rivilla2021a} were recently detected in the molecular cloud G+0.693-0.027  (hereafter referred to as G+0.693) and are considered to be key components in the formation of phospholipids. 

While its origin is not known, ethanolamine (hereafter EtA) alone accounts for 25\% of one of the two molecular subunits that form terrestrial phospholipids and has also been found in the Almahata Sitta meteorite \citep{Glavin_2010}. Here, its original formation might be ascribed to thermal decomposition of amino acids due to unusual conditions to which the parent asteroid might have been exposed. 
Grain-surface chemistry already showed that this prebiotic molecule can be formed through UV irradiation of interstellar ice analogues \citep{Ruaud_2015} and {\citet{rivilla2021a}} suggested that EtA might have formed through subsequent solid-phase hydrogenation steps.
{In this latter work, the authors mentioned that EtA formation might have involved the \ch{NH2CH2} radical, a result of \ch{NH2CH} hydrogenation. As extensively explained there, recent laboratory experiments \citep{Fedoseev_2015MNRAS, Ioppolo_2021NatAs} showed that the intermediate radicals \ch{NH2CH2} and hydroxymethyl \ch{CH2OH}, leading respectively to methylamine (\ch{CH3NH2}) and methanol (\ch{CH3OH}), are efficiently formed in the hydrogenation reactions and have been proposed as viable routes to the formation of  COMs \citet{Garrod_2006, Garrod_2008, Garrod_2013}. \ch{CH2OH} has been shown to play a significant role in the cold surface hydrogenation of CO molecules without involvement of UV- or cosmic-ray energetic processing of interstellar ices at 15 K, leading to glycolaldehyde \ch{(HC(O)CH2OH)}, ethylene glycol \ch{(H2C(OH)CH2OH)} \citet{Fedoseev_2015MNRAS}, and methyl formate \ch{(HC(O)OCH3)} \citet{Chuang_2016MNRAS}. Non-diffusive reactions between \ch{NH2CH2} and \ch{CH2OH} might therefore lead to EtA on dust grains \citet{rivilla2021a}. Additionally, EtA might also be formed through the hydrogenation of \ch{NH2CH2CO}, as a product of the reactions involving \ch{NH2CH2} and CO \citet{Suzuki_2018}.
On the other hand, in order to extend our comprehension of the chemical pathways leading to EtA and reported in Figure \ref{fig:EtA_form_process}, it would be helpful to also search the ISM for the intermediates highlighted in grey in that figure}, specifically in regions where they might migrate from the dust particles to the gas phase. Until now, it appears that the fewer atoms there are in a given molecule, the higher the probability of its detection in space \citep{McGuire_2022}. It therefore seems reasonable to begin an investigation of this proposed mechanism in ascending order of atoms involved. 

Referring to Figure \ref{fig:EtA_form_process}, \ch{HNCCO} (iminoethenone) might be formed through solid-phase N-addition to the ketenyl radical (HCCO) \citep{Charnley_2002}, which is formed in the gas-phase reaction CCH + OH \citep{Wakelam_2015}. Another route implies dust-grain surface reaction of ketene (\ch{H2CCO}) after two hydrogen abstractions, possibly reacting with the imine radical NH \citep{Rivilla_2021}. Nevertheless, \citet{Flammang_1994} argue that the probability of detecting HNCCO in the gas phase is low given its tendency to easily decompose into HNC + CO.
While this tendency to decompose might not be a problem in the ISM, it is in laboratory conditions, where the chemical production of \ch{HNCCO} needs to be sufficiently stable in order to allow characterisation of its rotational spectrum. 

Regarding \ch{NH2CH2CO}, the number of atoms, the presence of nitrogen, and the fact that it presents an open-shell electronic configuration further weaken the rationale for it to be a primary target for rotational spectroscopic studies.  Instead, the aldehyde analogue \ch{NH2CH2CHO}, which is a closed-shell molecule, has an additional hydrogen atom that confers a more pliable molecular structure, increasing the number of internal rotations and consequently decreasing the partition function and its line intensities. Therefore, our first reasonable target should be \ch{NH2CHCO}. 

Whether or not \ch{HNCCO} is produced in the condensed phase, it might in fact lead to \ch{NH2CHCO}, which is also proposed to be the product occurring between the barrierless reaction of \ch{NH3}, \ch{CO}, and atomic \ch{C} \citep{Krasnokutski_2021}. As the abundance of free carbon atoms is around half that of CO \citep{Tanaka_2011} in the  molecular clouds of the Galactic centre, this route could contribute to the formation of \ch{NH2CHCO} in G+0.693. Although three-body reactions are less efficient than two-body reactions  in the ISM, the fact that carbon atoms are expected to be highly reactive favours this scenario. The barrierless \ch{NH2CH} + CO reaction proposed by \citet{Suzuki_2018} might also contribute to the formation of \ch{NH2CHCO} in various astrophysical regions.

\begin{figure*}[tb!]
 \centering
 \includegraphics[width=1\textwidth]{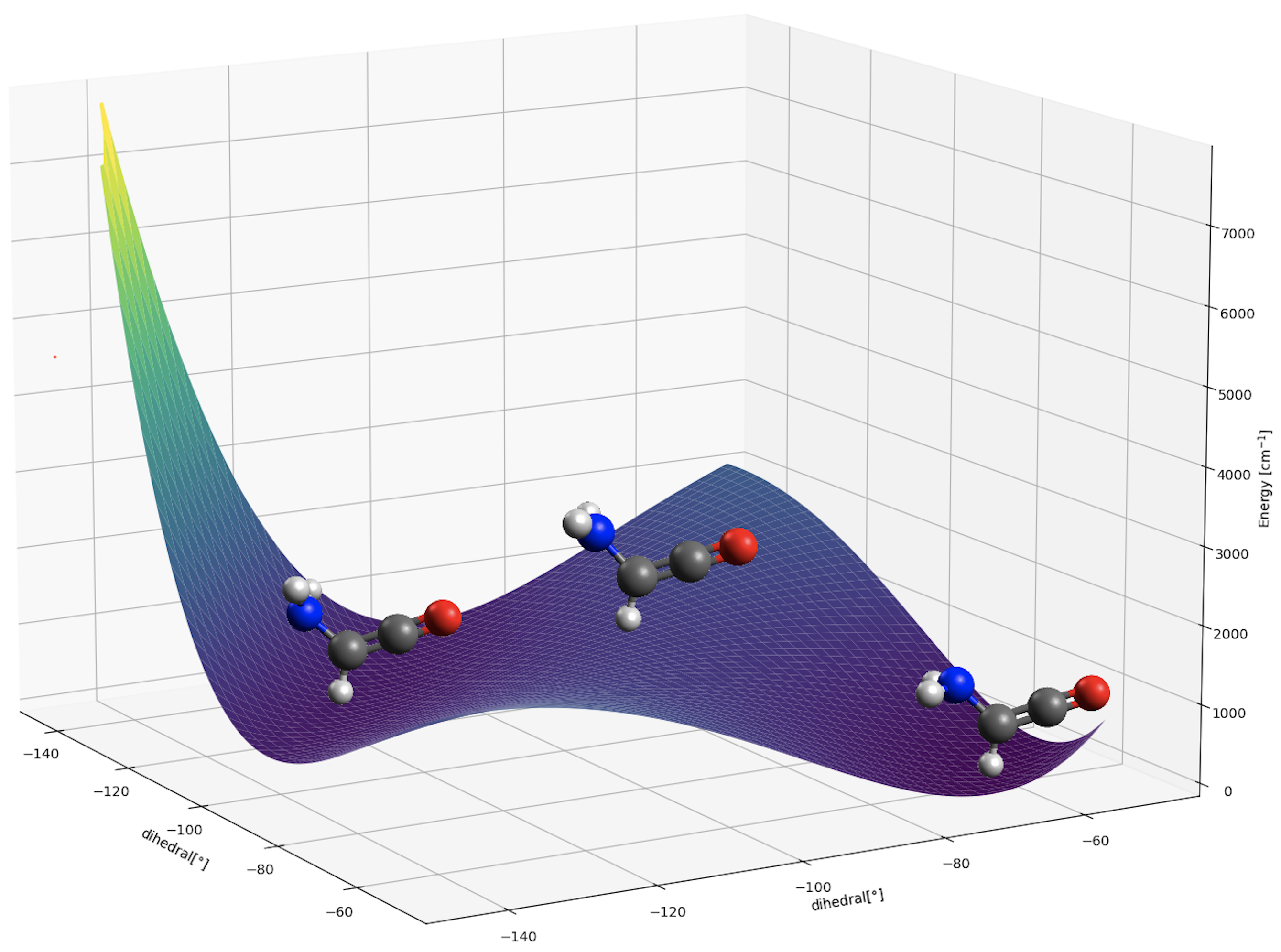}
 \caption{Potential energy surface scan of the \ch{NH2CHCO} conformer proposed by \citet{Krasnokutski_2021, Krasnokutski_2022} at the CCSD(T)-F12/cc-pCVTZ-F12 level of theory. Resolution is of two degrees. Dihedral values refer to the angles lying between the hydrogen of the primary amine and the CN bond.}
    \label{fig:PES}
\end{figure*}

In recent years, the number of amines detected in space has been rapidly increasing, undermining the idea that imines are the most abundant N-bearing molecules after cyanides. So far, in addition to EtA, methylamine \citep[\ce{CH3NH2},][]{Kaifu_1974}, hydroxylamine \citep[\ce{NH2OH},][] {rivilla2020b}, {and vynilamine \citep[\ce{C2H3NH2},][]{zeng2021} have been detected and ethylamine \citep[\ce{C2H5NH2},][]{zeng2021} has been tentatively detected.}
Among imines, methanimine \citep[\ce{CH2NH},][]{godfrey_1973}, E-cyanomethanimine \citep[\ce{HNCHCN},][]{Zaleski_2013}, prapargylimine \citep[\ce{HC3HNH},][]{bizzocchi2020propargylimine}, and ethanimine \citep[\ce{CH3CHCN},][]{Loomis_2013} have also been detected.  Ketenimine \citep[\ce{CH2CNH},][]{Lovas_2006}, {aziridine \citep[\ce{C2H5N},][]{Dickens_2001},} and allylimine \citep[\ce{CH2CHCHNH},][]{Alberton_2023a} have been tentatively detected, while an upper limit has been reported for propanimine \citep[\ce{CH3CH2CHNH},][]{Margules_2022}. In a recent paper, \citet{Krasnokutski_2022} tackled the formation pathway of \ch{NH2CHCO} and presented evidence for its formation in a 10K ice mixture in amine form. 

Nevertheless, the information obtained with the level of theory previously employed \citep{Krasnokutski_2022} can be improved with higher level calculations.
Additionally, this latter analysis (see also \citealt{Krasnokutski_2021}) does not explore the entire breadth of possible isomers for this molecule, most notably not reporting the iminic form.  While \ch{NH2CHCO} may be required for the reactions as given in Figure \ref{fig:EtA_form_process}, if this isomer easily breaks down into other forms, it will not be available to provide the chemistry as proposed.  Therefore, this present work explores \ch{NH2CHCO} isomers and also provides theoretical spectral data for their characterisation in the laboratory. This is the first step for possible analysis of their presence in space, providing insight into interstellar prebiotic chemistry.

\section{Computational details}\label{sec:exp}
\begin{figure*}[tb!]
 \centering
 \includegraphics[width=1.0\textwidth]{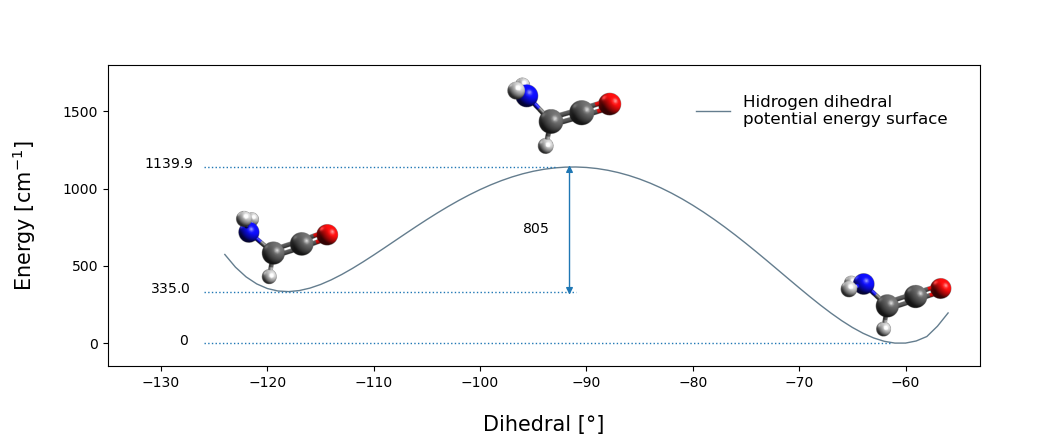}
 \caption{Energy profile of NH dihedral calculated at CCSD(T)-F12/cc-pCVTZ-F12 level of theory. The dihedral value refers to the angle lying between the hydrogen of the primary amine, and the CN bond. The second NH dihedral is kept constant with respect to the former.}
    \label{fig:erg_profile}
\end{figure*}

The theoretical computations performed in this work employ Gaussian09 \citep{frisch2009gaussian}, MOLPRO (2015.1 version)\footnote{For the current version of MOLPRO, see https://www.molpro.net.} \citep{werner2015molpro, MOLPRO_2020}, CFOUR (2.1-serial version\footnote{For the current version of CFOUR, see http://www.cfour.de.}) \citep{Matthews_2020}, and SPECTRO \citep{gaw1991spectro, Westbrook23}, this latter being a perturbational code to produce the final spectroscopic data. The scans of the potential energy surface (PES) of the \ch{NH2CHCO} isomers were performed employing Gaussian and MOLPRO. We employed the B3LYP \citep{Lee_DFT_1988, Becke_1993} hybrid functional theory and Møller–Plesset perturbation theory up to the second order \citep[MP2,][]{Moller_1934} in conjunction with the 6-311+G(d,p) basis set \citep{Andersson_6-311+G(dp)_2005} and the CCSD(T)-F12/UCCSD(T)-F12 level of theory \citep{Knizia_2009,Adler_2007} using the cc-pCVTZ-F12 basis set \citep{Hill_2010}. All isomer geometries are optimised with CCSD(T)-F12/cc-pCVTZ-F12. 

Spectroscopic constants are obtained using CFOUR and a MOLPRO-SPECTRO combination. CFOUR utilises the CCSD(T)/cc-pVXZ  with X = D, T level of theory; harmonic and anharmonic terms are computed employing second-order perturbation theory (VPT2) after the full cubic and the semidiagonal part of the quartic force field treatments, which in turn generate all vibrational constants apart from those of the form $\phi_{ijkl}$ \citep{Mills72, Watson-1977, Papousek82}. Additionally, the so-called F12-TcCR quartic force field (QFF) is employed to provide VPT2 results via SPECTRO \citep{Watrous22, Fortenberry22}.  This composite energy is defined at each displaced point from the core-including CCSD(T)-F12/cc-pCVTZ-F12 optimised geometry; CCSD(T)-F12/cc-pCVTZ-F12 energies (`TcC') are computed and combined with CCSD(T)/cc-pVTZ-DK relativistic effects (`R'), a method known to produce accuracies on the order of 1.0 cm$^{-1}$ compared to experiments \citep{Watrous22, Gardner21}. Subsequently, SPECTRO transforms the force constants from the QFF contributions to averaged and singly vibrationally excited principal rotational constants as well as the quartic and sextic centrifugal distortion constants \citep[][]{inostroza2011,Inostroza_2013}. 

For the specific case presented below, the rotational and distortion constants were obtained at CCSD(T)/cc-pXTZ level of theory only (with X = D, T). The CCSD(T)-F12-TcCR/cc-pCVTZ-F12 level was inferred using the scaling factors presented in Table \ref{tab:ab-initioadjusted} in Appendix \ref{sec:add-mat}. These scaling factors are obtained by computing the differences in rotational and distortion constants obtained at CCSD(T)-F12-TcCR/cc-pCVTZ-F12 and CCSD(T)/cc-pVTZ levels of theory for 10-\ch{HNCHCHO}. These differences are applied to the constants obtained for 9-\ch{HNCHCHO} at CCSD(T)/cc-pVTZ and comprise the effects given by the anharmonic corrections at the full QFF level and F12-TcCR contributions.
\indent\indent
\begin{figure*}[!]
 \centering
 \includegraphics[width=1.03\textwidth]{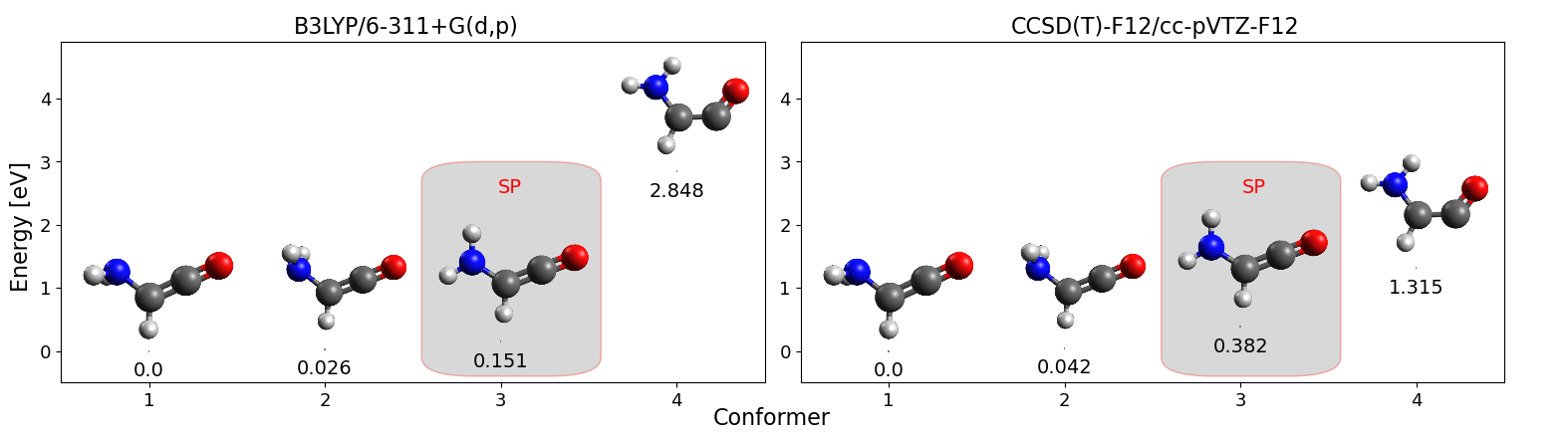}
 \caption{Comparison with the conformers proposed by \citet{Krasnokutski_2021}. On the left, at B3LYP/6-311+G(d,p) level of calculation, and on the right at the CCSD(T)-F12/cc-pCVTZ-F12 level. These conformers are the product of the C + \ch{NH3} + CO barrierless reaction. The SP connecting 2 and 3 \ch{NH2CHCO} conformers is reported in the grey box. For the isomer 4, the values were obtained employing UKS (spin-Unrestricted Kohn-Sham program) B3LYP/6-311+G(d,p) and UCCSD(T)-F12/cc-pCVTZ level of theory, for left and right plots, respectively.}
    \label{fig:erg_1}
\end{figure*}

\begin{table*}[!]
 \centering
 \caption{Energy of \ch{NH2CHCO} conformers herein found in respect to the work of Krasnokutski et al. 2021.} 
 \label{tab:erg_1}
 \smallskip
 \begin{tabular}{lr@{.}lr@{.}lr@{.}lcr@{.}l r@{.}l r@{.}l}
  \hline\hline \\[-1ex]
                                   & \mcl{6}{c}{Energy$^a$ [cm$^{-1}$]}          &              & \mcl{6}{c}{Energy$^a$ [eV]}                 \\[0.5ex]
  \cline{2-7} \cline{9-14}\\[-1.5ex]
                          &\mcl{2}{c}{B3LYP}       & \mcl{2}{c}{MP2}         &\mcl{2}{c}{CCSD(T)-F12}   &&    \mcl{2}{c}{B3LYP}    &      \mcl{2}{c}{MP2}     & \mcl{2}{c}{CCSD(T)-F12} \\[0.5ex]
      Conformers             &\mcl{2}{c}{6-311+G(d,p)}&\mcl{2}{c}{6-311+G(d,p)} & \mcl{2}{c}{cc-pCVTZ-F12} &&\mcl{2}{c}{6-311+G(d,p)} & \mcl{2}{c}{6-311+G(d,p)} & \mcl{2}{c}{cc-pCVTZ-F12}\\[0.5ex]
  \hline \\[-1.5ex]
  1                       &      0&0    &      0&0    &      0&0          &&      0&000    &    0&000     &      0&000     \\[0.5ex]
  2                       &    212&7    &    687&1    &    334&8          &&      0&026     &    0&085    &      0&042     \\[0.5ex]
  3$^b$                   &   1220&0    &   3578&7    &   3078&1          &&      0&151     &    0&444    &      0&382     \\[0.5ex]
  4                       &  22967&8$^c$&-142014&7$^d$&  10604&8$^e$  &&      2&848$^c$&  -17&607$^d$&      1&315$^e$ \\[0.5ex]
     \hline \\[-1.5ex]
\end{tabular}
\tablefoot{$^a$The energies are normalised to that of the lowest energy isomer 1. $^b$Saddle point connecting 2 and 3 \ch{NH2CHCO} conformers. $^{c,d,e}$Values obtained employing UKS (spin-Unrestricted Kohn-Sham program) B3LYP/6-311+G(d,p), UPM2/6-311+G(d,p), and UCCSD(T)-F12/cc-pCVTZ level of theory, respectively. The difference in the significant digits existing between values in Hartree and eV is maintained to justify the Hartree-eV conversion factor (1 a.u = 27.2114 eV). We note that the CO+NH\textsubscript{3}+C reaction lies approximately 1038, 1013, and 1060 eV above the 1-conformer for the three respective levels of theory.}
\end{table*}

\begin{figure*}[tb]
 \centering
 \includegraphics[width=1.0\textwidth]{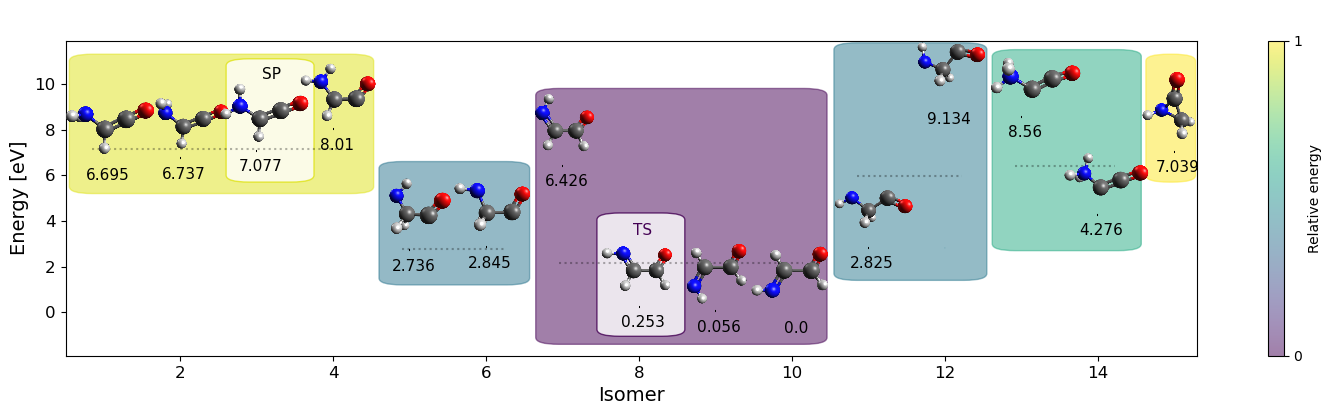}
 \caption{All new calculated conformers and isomers at CCSD(T)-F12/cc-pCVTZ-F12 level. Conformers of the same isomer are grouped in the same box. Each box is colour-coded according to the minimum energy of the relative conformer. From the most to the least favourable isomer, normalised colour scale ranges from violet to yellow, respectively. The SP and TS are both highlighted by lighter-coloured boxes. The average value of the isomer's conformers, which does not include TS and SP, and the values of each conformer are reported with a grey dotted line and a black solid line, respectively. The numerical value of the latter is also given.}
    \label{fig:isomers_CCSD(T)}
\end{figure*}

\section{Results and discussion}
A successful laboratory or even interstellar gas phase detection of a molecule is a direct consequence of the precise characterisation of the rotational constants {and the reaction kinetics involved}. These, in turn, rely upon reference data of some variety.  Typically, this is most readily provided by reliable quantum chemical computations as well as the identification of the lowest energy isomers of the molecular target \citep{Puzzarini_2023}.

\begin{table*}[tb]
 \centering
 \caption{Energy of all \ch{NH2CHCO} conformers and isomers.}
 \label{tab:erg_2}
 \smallskip
 \begin{tabular}{lr@{.}lr@{.}lr@{.}lcr@{.}l r@{.}l r@{.}l}
  \hline\hline \\[-1ex]
                                   & \mcl{6}{c}{Energy$^a$ [cm$^{-1}$]}       &              & \mcl{6}{c}{Energy$^a$ [eV]}                 \\[0.5ex]
  \cline{2-7} \cline{9-14}\\[-1.5ex]
                          &\mcl{2}{c}{B3LYP}       &\mcl{2}{c}{MP2}         &\mcl{2}{c}{CCSD(T)-F12} &&      \mcl{2}{c}{B3LYP}   &      \mcl{2}{c}{MP2}     & \mcl{2}{c}{CCSD(T)-F12} \\[0.5ex]
      Isomers             &\mcl{2}{c}{6-311+G(d,p)}&\mcl{2}{c}{6-311+G(d,p)}&\mcl{2}{c}{cc-pCVTZ-F12}&& \mcl{2}{c}{6-311+G(d,p)} & \mcl{2}{c}{6-311+G(d,p)} & \mcl{2}{c}{cc-pCVTZ-F12}\\[0.5ex]
  \hline \\[-1.5ex]
  1                       & -20726&1    & 129305&2    &  54003&2    &&   -2&567    &  16&032        &   6&695     \\[0.5ex]
  2                       & -20513&4    & 129992&2    &  54338&0    &&   -2&543    &  16&117        &   6&737     \\[0.5ex]
  3$^b$                   & -19506&0    & 132883&9    &  57081&3    &&   -2&418    &  16&475        &   7&077     \\[0.5ex]
  4                       &   2241&8    & -12709&6    &  64607&9    &&    0&278    &  -1&576        &   8&010     \\[0.5ex]
  5                       &   1521&5    & -15421&1    &  22066&0    &&    0&189     &  -1&912       &   2&736     \\[0.5ex]
  6                       &   1610&0$^c$&   8336&7$^d$&  22944&2$^e$&&    0&200$^c$&   1&034$^d$   &   2&845$^e$ \\[0.5ex]
  7                       &    802&3$^c$& -14423&3$^d$&  51829&9$^e$&&    0&099$^c$&  -1&788$^d$   &   6&426$^e$ \\[0.5ex]
  8$^f$                   &   2069&6$^c$& -13163&2$^d$&   2042&9$^e$&&    0&257$^c$&  -1&632$^d$   &   0&253$^e$ \\[0.5ex]
  9                       &    581&0    & -14524&0    &    450&8    &&    0&072     &  -1&801       &   0&056     \\[0.5ex]
 10                       &      0&0    &      0&0    &      0&0    &&    0&000     &   0&000       &   0&000     \\[0.5ex]
 11                       &   1471&5$^c$&   8276&0$^d$&  22782&5$^e$&&    0&182$^c$&   1&026$^d$   &   2&825$^e$ \\[0.5ex]
 12                       &   2737&6    &   8034&2    &  73671&8    &&    0&339     &   0&996       &   9&134     \\[0.5ex]
 13                       &  -5574&9    & 147548&4    &  69044&8    &&   -0&691     &  18&293       &   8&560     \\[0.5ex]
 14                       &  17097&2    &   2722&2    &  34490&7    &&    2&120     &   0&338       &   4&276     \\[0.5ex] 
 15                       & -19040&0    & 133322&8    &  56775&8    &&   -2&361     &  16&530       &   7&039     \\[0.5ex]
   \hline \\[-1.5ex]
\end{tabular}
\tablefoot{$^a$The energies are normalised to that of the lowest energy isomer, namely isomer 10. $^b$Saddle point connecting 2 and 3 \ch{NH2CHCO} conformers. $^{c,d,e}$Values obtained employing UKS (spin-Unrestricted Kohn-Sham program) B3LYP/6-311+G(d,p), UPM2/6-311+G(d,p), and UCCSD(T)-F12/cc-pCVTZ level of theory, respectively. $^f$Transition state connecting 7 and 9 \ch{HNCHCHO} conformers. The difference in the significant digits existing between values in Hartree [a.u.] and eV is maintained to justify the Hartree-eV conversion factor (1 a.u = 27.2114 eV). We note that the CO+NH\textsubscript{3}+C reaction lies approximately 1036, 1029, and 1066 eV above the 10-isomer, for the three respective levels of theory.}
\end{table*}

\subsection{Isomers and relative energies}\label{sec:isomer}

{There is a trend seen in the COMs detected in the ISM in that they are the energetically most favoured isomers. However, it is important to note that kinetics can also play a significant role in determining the outcome of chemical reactions in the ISM, even when the most stable isomer is not the kinetically most favoured product (e.g. \ch{CH2CCO}, \citealt{Shingledecker_2021}; cis- formic acid \ch{HCOOH}, \citealt{Cuadrado_2016}; and trans- methyl formate \ch{CH3OCHO}, \citealt{Neill_2012}). Consequently, it is possible to detect isomers with higher energies in the ISM, even if they are not the most stable isomers. In light of this, the isomer proposed by \citet{Krasnokutski_2021} and \citet{Krasnokutski_2022} could be detected in the ISM, even though it is less energetically favoured than others. Nevertheless, understanding the EtA chemical network requires the detection of the relevant species that might be involved. Relative energies are an important consideration, but the molecular precursors and energetic environment will greatly inform the products that are formed regardless of the relative energies.}

The first step in this work was to verify that {the energy associated with} the geometry of the \ch{NH2CHCO} conformer proposed by \citet{Krasnokutski_2022} is indeed the energetic minimum of this chemical formula. After geometrical optimisation of this same isomer at the CCSD(T)-F12/cc-pCVTZ-F12 level of theory, a potential energy surface (PES) scan was performed to study an energy profile of the isomer as a function of the two HNH dihedral angles. Figure \ref{fig:PES} reports the PES scan obtained through the scan at two-degree increments with CCSD(T)-F12/cc-pCVTZ-F12. The dihedral values refer to the angles lying between the hydrogen of the primary amine and the CN bond. Another isomer was determined to lie at a lower energy through this scan. This stability might be due to a more favourable steric arrangement of the amino hydrogens with respect to the nearby sp\textsuperscript{2} carbon.
Figure \ref{fig:erg_profile} shows the same analysis specifically along the coordinate for the equivalent hydrogens as the reflection of each other. The energy difference between these two conformers is clear at 335 cm\textsuperscript{-1}, and is equivalent to \(\sim \)480 K. The two conformers are separated by an 805 cm\textsuperscript{-1} (\(\sim \)1150 K) energy barrier, constituting an impediment to free hydrogen movement.

\begin{table*}[tbh!]
 \centering
 \caption{Theoretical frequencies associated with the vibration modes of 1- and 10-\ch{HNCHCHO} isomers.} 
 \label{tab:FOND_FREQ}
 \smallskip
 \begin{tabular}{l r@{.}l r@{.}l c r@{.}l r@{.}l}

  \hline\hline \\[-1ex]
  &\mcl{4}{c}{1-\ch{NH2CHCO}} && \mcl{4}{c}{10-\ch{HNCHCHO}}\\[0.5ex]
  \cline{2-5} \cline{7-10}\\[-1.5ex]
      ZPT$^a$ [cm$^{-1}$]& \mcl{4}{c}{10920.7}      &&  \mcl{4}{c}{10886.3}     \\[0.5ex]
      \hline\\[-1.5ex]
      \mcl{1}{l}{Mode} &\mcl{2}{c}{Harmonic [cm$^{-1}$]} & \mcl{2}{c}{Anharmonic$^b$ [cm$^{-1}]$} && \mcl{2}{c}{Harmonic [cm$^{-1}$]} & \mcl{2}{c}{Anharmonic$^b$ [cm$^{-1}$]} \\[0.5ex]
  \hline \\[-1.5ex]
  \ 1  & 3611&4 & 3435&8 && 3474&9 & 3304&8 \\[0.5ex]
  \ 2  & 3530&4 & 3374&7 && 3096&8 & 2939&6 \\[0.5ex]
  \ 3  & 3171&4 & 3036&0 && 2990&9 & 2837&7 \\[0.5ex]
  \ 4  & 2192&3 & 2143&4 && 1773&6 & 1747&8 \\[0.5ex]
  \ 5  & 1671&2 & 1622&0 && 1672&5 & 1638&7 \\[0.5ex]
  \ 6  & 1437&7 & 1402&3 && 1405&0 & 1371&4 \\[0.5ex]
  \ 7  & 1224&1 & 1188&5 && 1369&2 & 1339&3 \\[0.5ex]
  \ 8  & 1182&0 & 1154&4 && 1228&9 & 1205&4 \\[0.5ex]
  \ 9  & 1033&8 & 1006&7 && 1119&9 & 1134&7 \\[0.5ex]
  \ 10 &  821&2 &  748&2 && 1040&5 & 1017&1 \\[0.5ex]
  \ 11 &  656&4 &  641&3 && 1033&5 & 1029&2 \\[0.5ex]
  \ 12 &  599&0 &  582&3 &&  701&2 &  692&9 \\[0.5ex]
  \ 13 &  531&5 &  537&9 &&  582&6 &  579&9 \\[0.5ex]
  \ 14 &  254&9 &  249&8 &&  342&2 &  351&6 \\[0.5ex]
  \ 15 &  218&2 &  216&9 &&  149&9 &  178&8 \\[0.5ex]
  \hline \\[-1.5ex]
\end{tabular}
\tablefoot{
Frequencies are obtained at the CCSD(T)-F12-TcCR/cc-pCVTZ-F12 level of theory.
$^a$ZPT stands for zero point energy.
$^b$Anharmonic frequencies take into account Fermi-resonance corrections.
}
\end{table*}
\begin{table*}[tbh!]
 \centering
 \caption{Theoretical spectroscopic constants determined for 1- and 10-\ch{HNCHCHO} isomers.} 
 \label{tab:ab-initio_1}
 \smallskip
 \begin{tabular}{lr@{.}lr@{.}lr@{.}lcr@{.}l r@{.}l r@{.}l}

  \hline\hline \\[-1ex]
                            &        \mcl{5}{c}{1-\ch{NH2CHCO}}          &              & \mcl{7}{c}{10-\ch{HNCHCHO}}        \\[0.5ex]
  \cline{2-7} \cline{9-14}\\[-1.5ex]
      \mcl{1}{l}{Rotational}& \mcl{2}{c}{CCSD(T)$^a$} &\mcl{2}{c}{CCSD(T)$^a$} &\mcl{2}{c}{CCSD(T)-F12-TcCR$^b$} && \mcl{2}{c}{CCSD(T$^a$)} & \mcl{2}{c}{CCSD(T$^a$)} & \mcl{2}{c}{CCSD(T)-F12-TcCR$^b$} \\[0.5ex]
      \mcl{1}{l}{constants} &\mcl{2}{c}{cc-pVDZ} & \mcl{2}{c}{cc-pVTZ} & \mcl{2}{c}{cc-pCVTZ-F12} && \mcl{2}{c}{cc-pVDZ} & \mcl{2}{c}{cc-pVTZ} & \mcl{2}{c}{cc-pCVTZ-F12} \\[0.5ex]
  \hline \\[-1.5ex]
  \ \textit{A$_0$} \ \ \ /MHz &42829&8   &43900&3   &44201&2    &&51798&0    &52843&6    &53096&0      \\[0.5ex]
  \ \textit{B$_0$} \ \ \ /MHz & 4531&0   & 4592&4   & 4637&1   && 4638&9     & 4703&9    & 4749&0      \\[0.5ex]
  \ \textit{C$_0$} \ \ \ /MHz & 4183&2   & 4245&2   & 4286&8    && 4257&6    & 4319&7    & 4359&8      \\[0.5ex]
  \ $\Delta_J$\ \ \ \ /kHz    &    2&559 &    2&628 &    2&657 &&    0&987   &    1&024  &    1&041    \\[0.5ex]
  \ $\Delta_K$\ \ \ /MHz      &    2&569 &    2&797 &    2&801 &&    0&415   &    0&433  &    0&433 \\[0.5ex]
  \ $\Delta_{JK}$ \  /kHz     & -113&571 & -120&140 &    0&0   &&   -7&277   &   -7&174  &    0&0      \\[0.5ex]
  \ $\delta_J$ \ \  \ \ /Hz   &  523&470 &  536&605 &  542&934 &&  103&669   &  106&096  &  108&204    \\[0.5ex]
  \ $\delta_K$ \ \ \ /kHz     &   13&238 &   12&821 &   11&969 &&    5&111   &    5&300  &    5&401    \\[0.5ex]
  \ $\Phi_J$ \ \  \ /mHz      &    9&280 &   10&057 &   10&156 &&    0&196   &    0&200  &    0&204 \\[0.5ex]
  \ $\Phi_{JK}$ \ /mHz        &  -95&168 & -132&722 & -136&102 &&   -2&074   &   -2&376  &   -2&355   \\[0.5ex]
  \ $\Phi_{KJ}$ \  /Hz        &  -14&904 &  -15&973 &  -15&961 &&   -0&546   &   -0&563  &   -0&565 \\[0.5ex]
  \ $\Phi_{K}$ \ \ \ /Hz      &  464&261 &  529&158 &  527&549 &&    8&553   &    9&209  &    9&292    \\[0.5ex]
  \ $\phi_{J}$ \ \ \ \ /mHz   &    3&228 &    3&484 &    3&519 &&    0&072   &    0&074  &    0&076 \\[0.5ex]
  \ $\phi_{JK}$ \ \ /mHz      &  152&277 &  173&907 &  184&522 &&    2&942   &    2&963  &    2&978    \\[0.5ex]
  \ $\phi_{K}$ \ \ \ \ /Hz    &   21&900 &   23&634 &   22&920 &&    0&793   &    0&828  &    0&833 \\[0.5ex]
  \hline \\[-1.5ex]
\end{tabular}
\tablefoot{Geometries initially optimised at CCSD(T)-F12/cc-pCVTZ-F12 level of theory for the energy calculation were then optimised at the relative basis set.
$^a$Harmonic and anharmonic corrections were calculated using VPT2 with the F12-TcCR QFF.}
\end{table*}

While the studies mentioned earlier \citep{Krasnokutski_2021, Krasnokutski_2022} make use of two lower-level theoretical approaches, namely the B3LYP density functional theory (DFT) and MP2, both with the 6-311+G(d,p) Pople basis set, the present analysis of the isomers goes beyond this with coupled-cluster theory. In Table \ref{tab:erg_1}, the isomerisation energies obtained in this work are reported in cm\textsuperscript{-1} and eV. The results obtained at the same level of theory as the previous studies are flanked by the present ones from CCSD(T)-F12/cc-pCVTZ-F12. Figure \ref{fig:erg_1} shows a comparison between B3LYP/6-311+G(d,p) (previously used as reference in \citet{Krasnokutski_2021} and \citet{Krasnokutski_2022}) and CCSD(T)-F12/cc-pCVTZ-F12 levels of theory. The isomers are named starting from the lowest energy in cardinal numbers: 1 and 2 are the lowest- and highest-energy \ch{NH2CHCO} conformers, respectively; 3 and 4 refer to the saddle point (SP) that connects the previous conformer to the next one and an additional conformer found to be higher in energy, respectively.

The higher-level CCSD(T)-F12 energy difference between conformer 1 and 2 is double the amount suggested by DFT. 
The gap between conformer 1 and SP 3 passes from 0.1513 eV (\(\sim \)1775 K) in  B3LYP/6-311+G(d,p) to 0.3816 eV (\(\sim \)4428 K) in CCSD(T)-F12/cc-pCVTZ-F12; that is, it almost triples.  Differences between 1 and 4 go from 2.8476 eV (\(\sim \)33046 K) to 1.3148 eV (\(\sim \)15258 K), reducing by roughly a factor of one-half when explicitly correlated coupled-cluster theory is utilised. 

\begin{figure*}[tb!]
 \centering
 \includegraphics[width=1\textwidth]{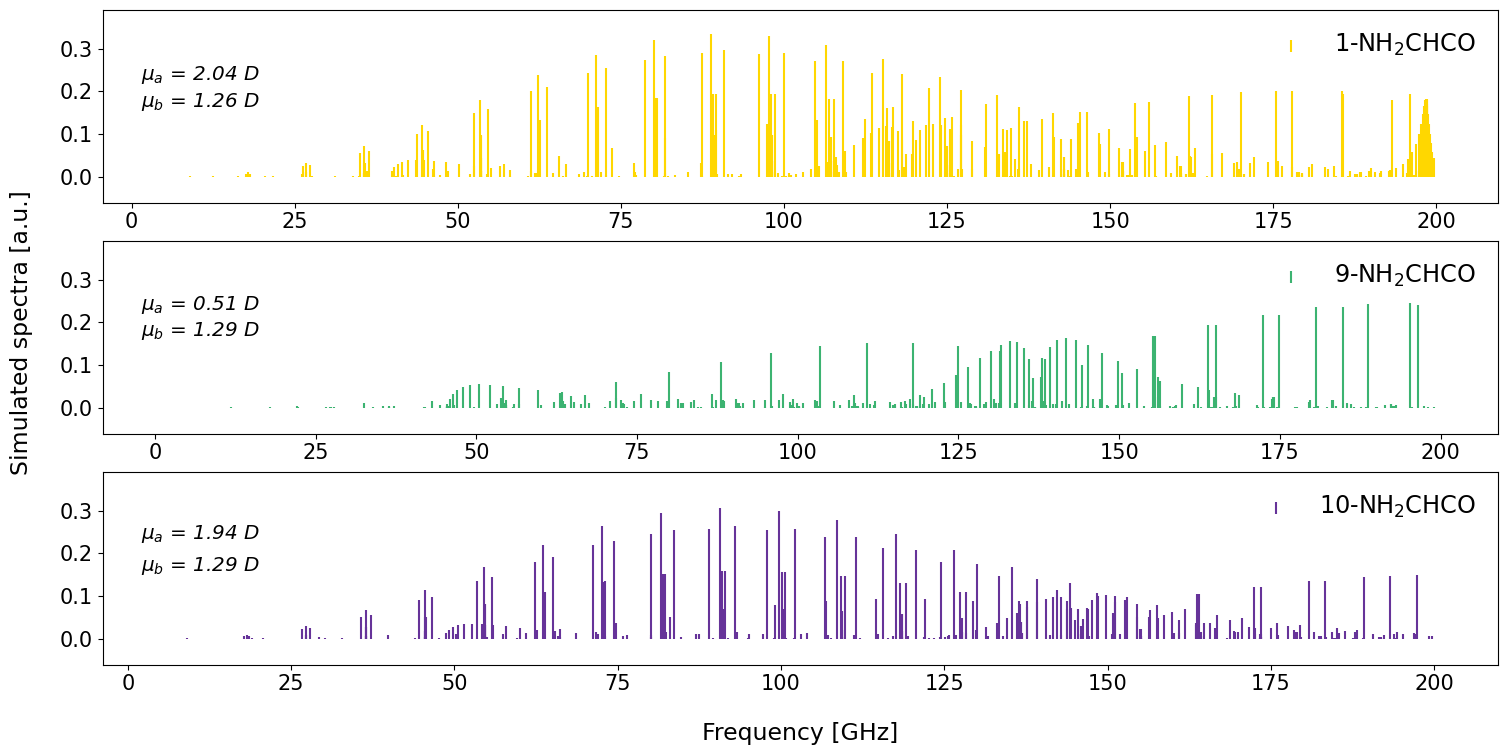}
 \caption{Simulated spectra at 15K of 1-, 9-, and 10-\ch{HNCHCHO} employing the rotational constants presented in Tables \ref{tab:ab-initio_1} and \ref{tab:ab-initio_isomer_9}. Theoretical spectra are produced using the highest level of calculation for each isomer, namely CCSD(T)-F12-TcCR QFF for 10- and 1-\ch{NH2CHCO} and the scaled CCSD(T)-F12-TcCR QFF for 9-\ch{HNCHCHO} (for which more details are provided in Table \ref{tab:ab-initioadjusted} of Appendix \ref{sec:add-mat}).}
    \label{fig:spectra}
\end{figure*}

\begin{table*}[tb!]
 \centering
 \caption{Theoretical spectroscopic constants determined for 9-\ch{HNCHCHO} isomer.}
 \label{tab:ab-initio_isomer_9}
 \smallskip
 \begin{tabular}{l r@{.}l  r@{.}l r@{.}l}

  \hline\hline \\[-1ex]
                            & \mcl{6}{c}{9-\ch{HNCHCHO}}        \\[0.5ex]
  \cline{2-7}\\[-1.5ex]
      \mcl{1}{l}{Rotational}& \mcl{2}{c}{CCSD(T)$^a$} & \mcl{2}{c}{CCSD(T)$^a$} & \mcl{2}{c}{CCSD(T)-F12-TcCR$^b$} \\[0.5ex]
      \mcl{1}{l}{constants} & \mcl{2}{c}{cc-pVDZ} & \mcl{2}{c}{cc-pVTZ} & \mcl{2}{c}{cc-pCVTZ-F12} \\[0.5ex]
  \hline \\[-1.5ex]
  \ \textit{A$_0$} \ \ \ /MHz& 49363&3    &50035&8    &50288&2   \\[0.5ex]
  \ \textit{B$_0$} \ \ \ /MHz&  4628&3    & 4707&1    & 4752&3   \\[0.5ex]
  \ \textit{C$_0$} \ \ \ /MHz&  4231&8    & 4303&2    & 4343&2   \\[0.5ex]
  \ $\Delta_J$\  \ \ /kHz    &     1&043  &    1&088  &    1&105   \\[0.5ex]
  \ $\Delta_K$ \ \ /MHz      &     0&361  &    0&372  &    0&372   \\[0.5ex]
  \ $\Delta_{JK}$ \ /kHz     &    -6&978  &   -6&800  &   -6&800   \\[0.5ex]
  \ $\delta_J$ \ \ \ \ /Hz   &   114&123  &  118&658  &  120&766    \\[0.5ex]
  \ $\delta_K$ \ \ \ /kHz    &     5&354  &    5&629  &    5&730   \\[0.5ex]
  \ $\Phi_J$ \ \ \ /mHz      &     0&192  &    0&177  &    0&181   \\[0.5ex]
  \ $\Phi_{JK}$ \ /mHz       &     1&160  &    2&524  &    2&503    \\[0.5ex]
  \ $\Phi_{KJ}$ \  /Hz       &    -0&623  &   -0&681  &   -0&683   \\[0.5ex]
  \ $\Phi_{K}$ \ \ \ /Hz     &     7&346  &    7&784  &    7&867   \\[0.5ex]
  \ $\phi_{J}$ \ \ \ \ /mHz  &     0&076  &    0&075  &    0&077   \\[0.5ex]
  \ $\phi_{JK}$ \ \ /mHz     &     3&525  &    3&644  &    3&658    \\[0.5ex]
  \ $\phi_{K}$ \ \ \ \ /Hz   &     0&808  &    0&857  &    0&861   \\[0.5ex]
  \hline \\[-1.5ex]
\end{tabular}
\tablefoot{Geometries initially optimised at CCSD(T)-F12/cc-pCVTZ-F12 level of theory for the energy calculation were then optimised at the relative basis set.
$^a$Harmonic and anharmonic corrections were calculated from VPT2 via the F12-TcCR QFF.}
\end{table*}
We conducted a wider PES isomer
analysis in an attempt to understand whether the two \ch{NH2CHCO} conformers identified so far are effectively the most energetically favoured. For consistency, our analysis used the B3LYP/6-311+G(d,p), MP2/6-311+G(d,p), and CCSD(T)-F12/cc-pCVTZ-F12 levels of theory; the results are summarised in Table \ref{tab:erg_2}. The values are reported both in cm\textsuperscript{-1} and eV and are normalised with respect to the new energy minimum obtained at the highest level calculation. Figure \ref{fig:isomers_CCSD(T)} reports the isomerisation energies at the CCSD(T)-F12/cc-pCVTZ-F12 level of theory, which better constrain the energies of \ch{NH2CHCO} isomers and conformers \citep{Adler_2007,Rauhut_2009, Knizia_2009}. The molecules are grouped into distinct isomeric families. Within each of these, the analysed conformers are shown. The isomer families are colour-coded by taking the energy of the lowest-energy isomer as a reference, giving a total of six distinct isomers. Saddle points and transition states (TSs) are marked with lighter coloured boxes. This results in six distinct groups with colours ranging from yellow to violet for the highest to the lowest energies, respectively.

In Table \ref{tab:erg_2}, all of the isomers and conformers are normalised to the new minimum obtained at CCSD(T)-F12/cc-pCVTZ-F12 level of theory. The grey dotted line refers to the average value of the energy of the conformers in the relative coloured box \footnote{SPs and TSs are not considered.}, while the black solid line shows the energy value of the conformers themselves. The names assigned to the additional isomers do not take into account the ascending order of energy but instead refer to the order in which they were included in the analysis. The conformers calculated above were the least energetically favourable  after the
cyclic isomer. The iminic isomer (violet box) is the lowest-energy isomer, making it a primary target for spectroscopic exploration. 

In contrast to two previous studies \citep{Redondo_2018, Drabkin_2023}, we identified  three different minima for this iminoacetaldehyde, that in turn can be recognised as a tautomer of aminoketene. Excluding the highest-energy 7-\ch{HNCHCHO} and the TS (considered  a minimum in \citet{Redondo_2018}), conformers 9- and 10-\ch{HNCHCHO} lie fairly close in energy. The latter is the lowest-energy form of this molecular class but is only 0.056 eV (\(\sim \)650 K) lower than 9-\ch{HNCHCHO}, which differs only in the spatial arrangement of the imine hydrogen. This difference was estimated to be of 1.12 eV (\(\sim \)13000 K) at CCSD(T)/aug-cc-pVQZ//CCSD/cc-pVTZ level of theory in \citet{Redondo_2018}. 7-\ch{HNCHCHO} and 8-\ch{HNCHCHO} (TS), in turn, are analogues of the 9 and 10 conformers, the difference being that they have the iminic nitrogen on the same side as the aldehyde group. The TS is 0.253 eV (\(\sim \)2936 K) above, while 7-\ch{HNCHCHO} lies higher in energy (6.426 eV, or \(\sim \) 74500K). 
Isomers 13 and 14, which are enclosed in the green box, are at a higher energy level than the global minimum. Because the N atom is formally positively charged and the adjacent C atom is negatively charged, the system can be stabilised by proton transfer to produce aminoketene as well \citep{Krasnokutski_2022}. Thus, the proton transfer can occur from the \ce{NH3-group} present in isomer 13 to the adjacent negative carbon to lead to the formation of isomer 1  by an exothermic process (-1.8 eV). On the other hand, in the case of isomer 14, the proton transfer is from the \ce{NH3-group} of isomer 14 to the carbon to form isomer 1. This is an endothermic process with an energy barrier of about 2.4 eV.
In the previous studies \citep{Redondo_2018, Drabkin_2023},  the energies obtained at CCSD(T)/aug-cc-pVQZ//CCSD/cc-pVTZ level for the imine isomers are comparable with those obtained in this study at the B3LYP/6-311+G(d,p) level. In the former case, isomers 9, 8, and 7 have energies normalised to isomer 10 of 0.052 eV, 0.239 eV, and 0.085 eV, respectively. In the latter case, values are of 0.072 eV, 0.257 eV, and 0.099 eV for isomers 9, 8, and 7, respectively (Table \ref{tab:erg_2}). Interestingly, the energy value of isomer 7 differs substantially from the trends mentioned above, resulting in it lying much higher in energy than the other imines when the CCSD(T)-F12 method is employed. This difference has to be ascribed to the theoretical F12 treatment, whereby explicitly correlated effects are taken into account. This method includes the electron--electron interaction at all distances, rather than approximating it at long distances, allowing for improvement of the accuracy, especially for large molecules. Table \ref{tab:erg_2} shows how the F12 treatment effects not only isomer 7 but also many others.
This energy difference in isomer 7 is comparable to that found between the 10-\ch{HNCHCHO} and 1-\ch{NH2CHCO} isomers at the B3LYP/6-311+G(d,p) level, namely of 6.695 eV. For a comparison with the level of theory employed in the previous studies, Figure \ref{fig:isomers_B3LYP} in Appendix \ref{sec:b3lyp} shows the drastic difference in energy that these same isomers exhibit with B3LYP/6-311+G(d,p).

\subsection{Spectroscopic data}\label{sec:spect}

For interstellar identification purposes of this molecule, conformers 10-\ch{HNCHCHO} and 9-\ch{HNCHCHO} should be regarded as natural candidates for observations. This is in line with what \citet{Redondo_2018} suggested, but these authors report rotational and distortion constants  at CCSD/cc-pVTZ level of theory. This same idea is supported by the recent work of \citet{Drabkin_2023}, in which the authors were able to characterise the IR and UV-VIS spectrum of 10-\ch{HNCHCHO} and 9-\ch{HNCHCHO} employing a deposition of 2-azidoacetaldehyde as a precursor on a large excess of argon onto the cold matrix window. In order to be able to detect these molecules in molecular clouds, the next step would then be the characterisation of such molecules in the gas phase. In the present work, we provide state-of-the-art quantum chemical calculations in order to better constrain the rotational and distortion constants of these promising candidates. 

Both of the isomers are closed-shell molecules, easing the computational rotational analysis propaedeutic to the experimental characterisation of the rotational constants. The same cannot be said of isomers 5 and 6 as these are open-shell structures. Starting from the optimised geometry obtained at the CCSD(T)-F12/cc-pCVTZ-F12 level of theory, we computed the F12-TcCR QFF for isomers 1 and 10. Fundamental harmonic frequencies and Fermi-resonance corrected frequencies are reported in Table \ref{tab:FOND_FREQ}. Zero-point energy values are also provided.

We obtained sets of rotational and distortion constants at CCSD(T)/cc-pVXZ (X = D, T) and F12-TcCR levels of theory. Rotational and distortion constants of these two isomers are reported in Table \ref{tab:ab-initio_1} using the Watson A-reduced Hamiltonian.
The choice of this Hamiltonian was based on parameters $R_5$ and $R_6$ reported in Table \ref{tab:dipole}.Please check that I have retained your intended meaning. The latter are used to determine the free parameters $s_{111}(S)$ and $s_{111}(A)$, whose formulae are given in the equations \ref{eq:s111(S)} and \ref{eq:s111(A)}, respectively. $B_z$, $B_x$ and $B_y$, in turn, correspond to the rotational constants $A_0$, $B_0$, and $C_0$.

\begin{equation} \label{eq:s111(S)}
    s_{111}(S) = \frac{2R_5}{2B_z-B_x-B_y}
,\end{equation}

\begin{equation} \label{eq:s111(A)}
    s_{111}(A) = -\frac{4R_6}{B_x-B_y}
.\end{equation}

The free parameter $s_{111}$ takes part in the unitary transformation of the Hamiltonian, and as suggested by \citet{watson_1978}, its value should be as low as possible. Historically, the S and A reductions were conceived to approach near-symmetric prolate or oblate rotors ($\kappa = -1/+1$) and asymmetric rotors ($\kappa = 0$), respectively. However, the increase in computational power in recent decades and the minimal differences found here in the $s_{111}$ values mean that it is not strictly necessary to adopt one reduction over the other. For these reasons, as the values of $s_{111}(A)$ and $s_{111}(S)$ are both of the order of $10^{-7}$ or less, we adopted the Watson A-reduced Hamiltonian.

For isomer 9, the rotational and distortion constants are reported in Table \ref{tab:ab-initio_isomer_9}, making use of the scaling factors presented in Table \ref{tab:ab-initioadjusted} in Appendix \ref{sec:add-mat}. This strategy was adopted in order to overcome the low isomerisation barrier existing between isomers 9 and 10, which might have prevented a VPT2 analysis.

The different geometrical configuration of the isomers analysed in this section is clear when comparing the rotational constants. Taking the highest level of theory as a reference, 10- and 9-\ch{HNCHCHO} have similar \textit{A$_0$, B$_0$}, and \textit{C$_0$} rotational constants from which 1-\ch{NH2CHCO} clearly deviates. Figure \ref{fig:spectra} shows the synthetic spectra obtained by computing the rotational and distortion constants at 15K, an excitation temperature (T$_{ex}$) found for most of the molecules in the G+0.693 molecular cloud (see e.g. \cite{Zeng_2018}). At 15K, the Planck function peaks at around 90 GHz for 10-\ch{HNCHCHO} and 1-\ch{NH2CHCO}, and at around 200 GHz for 9-\ch{HNCHCHO}. The values of the dipole moments are $\mu_a = 2.04$ D, $0.51$ D, and $1.94$ D and $\mu_b = 1.26$ D, $1.29$ D, and $1.29$ D for 1-\ch{NH2CHCO} and 9- and 10-\ch{HNCHCHO}, respectively. These values are provided in Table \ref{tab:dipole}. As a consequence, the spectra of 10-\ch{HNCHCHO} and 1-\ch{NH2CHCO} are dominated by \textit{a}-type transitions, and that of 9-\ch{HNCHCHO} is dominated by \textit{b}-types. 

The fundamental \textit{a}-type transition $J_{K_a,K_c} = 1_{0,1}-0_{0,0}$ is given by the sum of the two rotational constants \textit{B$_0$} and \textit{C$_0$}. For assignment purposes, 1-\ch{NH2CHCO} and 10-\ch{HNCHCHO} have fundamental frequencies at 9.108 GHz and 8.923 GHz, respectively. In the case of 9-\ch{HNCHCHO} , the \textit{b}-type fundamental rotational transition $J_{K_a,K_c} = 1_{1,1}-0_{0,0}$ is instead given by the sum of \textit{A$_0$} and \textit{C$_0$} and falls at 4.594 GHz. In this case, all the isomers are prolate molecules, implying that the biggest rotational constant is \textit{A$_0$}. As a result, the uncertainty over the \textit{a}-type fundamental transitions will be significantly reduced compared to the \textit{b}-type fundamental transition. 
Compared to previous studies \citep{Redondo_2018, Drabkin_2023}, the higher level of theory employed here includes the additional corrections of core electron correlation, explicit correlation along with its faster basis set convergence, scalar relativity, and the Coriolis resonances in addition to Fermi resonance polyads, and provides a whole set of new sextic distortion constants.

\begin{table}[tb!]
 \centering
 \caption{Vibrationally averaged dipole moments of 1-\ch{NH2CHCO}, 9- and 10-\ch{HNCHCHO} isomers.}
 \label{tab:dipole}
 \smallskip
 \begin{tabular}{l c c r@{.}l r@{.}l r@{.}l}

  \hline\hline \\[-1ex]
                              &&& \mcl{6}{c}{\ch{Isomer}}        \\[0.5ex]
  \cline{4-9}\\[-1.5ex]
  \mcl{1}{l}{Parameter}& \mcl{1}{l}{Unit} && \mcl{2}{c}{1-} & \mcl{2}{c}{9-} & \mcl{2}{c}{10-} \\[0.5ex]
  \hline \\[-1.5ex]
   $R_5$                     &KHz&&    6&64       &   -0&71 &-0&58 \\
   $R_6$                     &Hz&&    -0&29       &   -4&62 &-4&05 \\
   $\mu_a$                     &D&&     2&04      &    0&51 & 1&94 \\
   $\mu_b$                     &D&&     1&26      &    1&29 & 0&98 \\
   $\mu_c$                     &D&&     0&0       &    0&0  & 0&0  \\
   $\kappa^a$                  & &&    -0&98      &   -0&98 &-0&98 \\

  \hline \\[-1.5ex]
\end{tabular}
\tablefoot{$^a\kappa$ is the asymmetric rotor parameter given by $(2B-A-C)/(A-C)$ \citep{Gordy_1984}. $\mu_x$ is the dipole moment with respect to the $x$ inertia axis. }
\end{table}

From a laboratory point of view, product formation being equal, 10-\ch{HNCHCHO} remains the most promising isomer; not only because it is the most energetically favoured, but also because it should exhibit the most intense transitions.
It therefore serves as the best observational target for this isomer family. Albeit to  a lesser extent, isomer 9 remains a valid alternative, even if it presents critical problems. While it exhibits the second-most energetically favoured structure, it does not present lines as  intense as those of 10-\ch{HNCHCHO}  at low frequencies (please refer to Figure \ref{fig:spectra}). Instead, isomer 1, although exhibiting the strongest transitions, is only the ninth of the isomers analysed in this work in order of energy, making searches for it in space more complicated and its detection less straightforward.


\section{Conclusions}\label{sec:obs}

Synthetic spectra computed employing very high-level ab initio calculation support the idea that 10-\ch{HNCHCHO}  is the most promising candidate for ISM observations among the molecules investigated here. 10-\ch{HNCHCHO} is one of the isomers of \ch{NH2CHCO} currently thought to be a crucial step in the formation route leading to EtA, a prebiotic molecule recently detected in space {\citep{rivilla2021a}}). The chemical pathway leading to its formation is shown in Figure \ref{fig:EtA_form_process}. 10-\ch{HNCHCHO} might help to elucidate 
 not
only the chemical network involving EtA, but also other chemical routes of astrochemical relevance, broadening our understanding of the chemistry and physics of the ISM.

In this paper, expanding on the contributions of previous work \citep{Krasnokutski_2021, Krasnokutski_2022}, we extended the study of \ch{NH2CHCO} to its different isomeric forms.
The isomer previously thought to be the lowest in energy and experimentally observed in a condensed-phase matrix \citep{Krasnokutski_2021, Krasnokutski_2022} is actually only the ninth most energetic out of 15 structural and conformational isomers that might be present in the gas-phase within the ISM. Of these 15 isomers, 13 were studied in this work; one is found to be a SP and another a TS, respectively. In contrast to previously reported findings, the most energetically favourable isomer of \ch{NH2CHCO} is not in the amine but rather the imine form \ch{HNCHCHO}. 

In contrast to what was observed by \citet{Redondo_2018} and \citet{Drabkin_2023}, the latter includes three minima, two of which in particular exhibit the lowest isomer or conformational energies. The rotational and distortion constants of the two lowest-energy isomers 9-\ch{HNCHCHO} and 10-\ch{HNCHCHO}, correctly identified as such in the previous studies \citep{Redondo_2018,Drabkin_2023}, were herein obtained at a higher level of theory, taking into account core electron correlation, explicit correlation along with its faster basis set convergence, scalar relativity, and VPT2 resonances. These new results should help to guide the community towards spectroscopic analysis of and observational searches for the above-mentioned species, allowing us to determine whether or not and in what quantities these molecules might take part in the processes that ultimately lead to the
 emergence of life.

        
        \begin{acknowledgements}
        We are grateful for the referee's assistance in making our paper a better contribution to the field. We thank the Max Planck Society for the financial support. NI acknowledges support from PCI-ANID Grant REDES190113 and from FONDECYT Grant 1241193. RCF acknowledges support from the University of Mississippi's College of Liberal Arts and NASA Grant 22-A22ISFM-0009.
        
        \end{acknowledgements}
        
        
        \bibliographystyle{aa}
        \bibliography{biblio}
        

\appendix

\section{B3LYP/6-311+G(d,p) energy diagram} \label{sec:b3lyp}
\begin{figure*}[tb!]
\centering
 \includegraphics[width=1.0\textwidth]{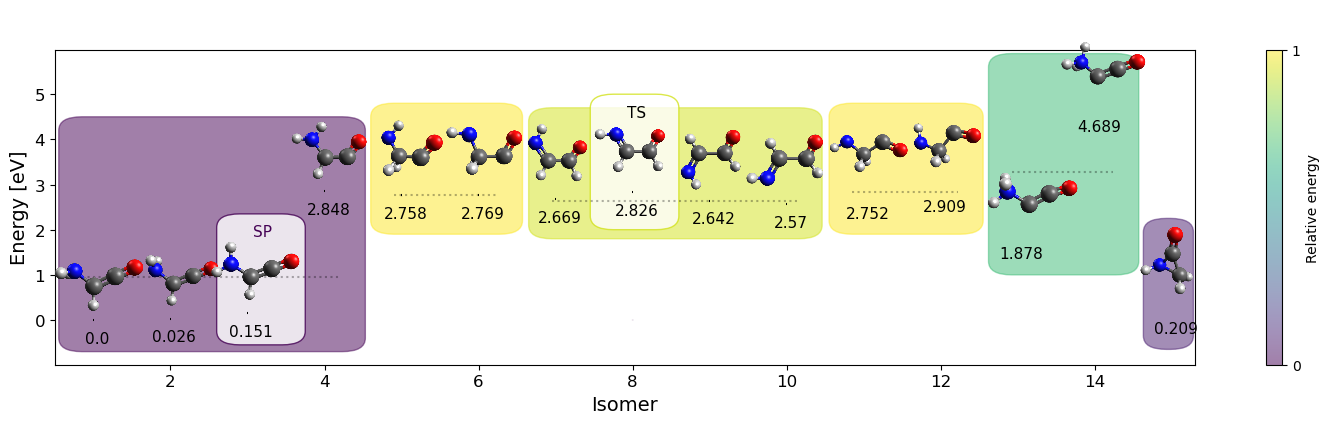}
 \caption{All new calculated conformers and isomers at B3LYP/6-311+G(d,p) level, normalised to the 1-\ch{NH2CHCO} minimum. Conformers of the same isomer are grouped in the same box. Each box is colour-coded according to the minimum energy of the relative conformer. From the most to the least favourable isomer, normalised colour scale ranges from violet to yellow, respectively. SP (saddle point) and TS (transition state) are both highlighted by lighter-coloured boxes. The average value of the isomer's conformers, which does not include TS and SP, and the values of each conformer are reported with the grey dotted line and the black solid line, respectively. The numerical value of the latter is also given.}
    \label{fig:isomers_B3LYP}
\end{figure*}

To have a direct comparison with Figure \ref{fig:isomers_CCSD(T)}, in Figure \ref{fig:isomers_B3LYP} we provide the results obtained at the B3LYP level of theory. We note how at this level of calculation the energy order of the isomer families is completely reversed. Amines are found to be the most energetically favoured, while in turn imines are considerably high in energy.

\section{Details of the theoretical calculations} \label{sec:add-mat}

\indent\indent
We obtained the rotational and distortion constants of 9-\ch{HNCHCHO}  using two approaches. The first comprises the CCSD(T)/cc-pVXZ level of theory, with X = D and T, 
for which harmonics and anharmonics terms were calculated employing the second-order perturbation theory (VPT2) after the full cubic and the semidiagonal part of the quartic force field treatments, that in turn generate all vibrational constants except those in the form $\phi_{ijkl}$ \citep{Mills72, Watson-1977, Papousek82}. The second involved the use of a scaling factor obtained from the differences in the values of the CCSD(T)-F12-TcCR/cc-pCVTZ-F12 and CCSD(T)/cc-pVTZ level of theory constants for 10-\ch{HNCHCHO}, to be applied at constants obtained at the highest 9-\ch{HNCHCHO} level of theory. These scaling factors comprise the effects given by the anharmonic corrections at the full quartic force field (QFF) level, taking into account F12-Triples Consistent Correlation analysis, core electron
Correlation (cC), and relativistic effect (R) contributions (F12-TcCR contributions). Also, VPT2 stands for Vibrational Perturbation Theory to second order force fields (VPT2). Scaling factors obtained from Isomer 1 were not employed because of its geometrically dissimilarity in comparison with isomer 10, which in turn is the closest conformer. Table \ref{tab:ab-initioadjusted} reports the scaling factors obtained from 1- and 10-\ch{HNCHCHO}, and the rotational and distortion constants obtained for 9-\ch{HNCHCHO}.

\begin{table}[tbh!]
 \centering
 \caption{Theoretical scaling factors obtained from 1- and 10-\ch{HNCHCHO} isomers}
 \label{tab:ab-initioadjusted}
 \smallskip
 \begin{tabular}{l r@{.}l c r@{.}l c r@{.}l r@{.}l r@{.}l}

  \hline\hline \\[-1ex]
                            &        \mcl{5}{c}{Diff. relative to \ch{NH2CHCO} isomers}
                            \\[0.5ex]
  \cline{2-6}\\[-1.5ex]
      \mcl{1}{l}{Rotational}& \mcl{5}{c}{$\Delta^a$}      \\[0.5ex]
      \mcl{1}{l}{constants} &\mcl{2}{c}{1-\ch{NH2CHCO}} & \mcl{3}{c}{10-\ch{HNCHCHO}} \\[0.5ex]
  \hline \\[-1.5ex]
  \ A$_0$ \ \ \        /MHz  & 300&9    && 252&4   \\[0.5ex]
  \ B$_0$ \ \ \        /MHz  &  44&7    &&  45&1   \\[0.5ex]
  \ C$_0$ \ \ \        /MHz  &  41&6    &&  40&0   \\[0.5ex]
  \ $\Delta_J$ \ \ \ /kHz    &   0&029  &&  0&017  \\[0.5ex]
  \ $\Delta_K$ \ \ /MHz      &   0&004  && -0&000  \\[0.5ex]
  \ $\Delta_{JK}$ \ /kHz     &   0&0$^b$&&  0&0$^b$\\[0.5ex]
  \ $\delta_J$ \ \ \ \ /Hz   &   6&329  &&  2&108  \\[0.5ex]
  \ $\delta_K$ \ \ \ /kHz    &  -0&852  &&  0&101  \\[0.5ex]
  \ $\Phi_J$ \ \ \ /mHz      &   0&099  &&  0&004  \\[0.5ex]
  \ $\Phi_{JK}$ \ /mHz       &   3&380  && -0&021  \\[0.5ex]
  \ $\Phi_{KJ}$ \  /Hz       &  -0&012  && -0&002  \\[0.5ex]
  \ $\Phi_{K}$ \ \ \ /Hz     &  -1&609  &&  0&083  \\[0.5ex]
  \ $\phi_{J}$ \ \ \ \ /mHz  &   0&035  &&  0&002  \\[0.5ex]
  \ $\phi_{JK}$ \ \ /mHz     &  10&614  &&  0&015  \\[0.5ex]
  \ $\phi_{K}$ \ \ \ \ /Hz   &  -0&714  &&  0&005  \\[0.5ex]
  \hline \\[-1.5ex]
\end{tabular}
\tablefoot{The scaling factors were obtained by computing the difference between the constants calculated at CCSD(T)-F12/cc-pCVTZ-F12 QFF F12-TcCR level of theory and those calculated at CCSD(T)/cc-pVTZ VPT2 level of theory. Values are given for the isomer 1- and 10-\ch{HNCHCHO}. 
$^a$$\Delta$ refers to  the difference found for the constant concerned between CCSD(T)-F12-TcCR/cc-pCVTZ-F12 and CCSD(T)/cc-pVTZ level of theories.
$^b$For $\Delta_{JK}$ differences, values were not considered due to the values obtained for both isomers in table \ref{tab:ab-initio_1}.}
\end{table}

\end{document}